# SURFACE PLASMON-POLARITON PULSES
# IN THE FORM OF BRIGHT AND DARK SOLITONS

**Igor V. Dzedolik, Olga Karakchieva**

*Taurida National V. I. Vernadsky University,*
*4, Vernadsky Avenue,*
*95007, Simferopol, Ukraine,*
*dzedolik@crimea.edu , karakchieva@crimea.edu*

We consider the linear and nonlinear models of generation of the surface plasmon-polaritons on the boundary of nonmagnetic dielectric medium and nonmagnetic metal. We have shown how the three-dimensional incident wave is transformed to the fluxes of surface plasmon-polaritons at the first and second harmonics of the TM-mode. These "slow" and "fast" fluxes of the surface plasmon-polaritons are formed at the both harmonics at the weak interaction. We obtain that the pulses of surface plasmon-polaritons propagate as the bright and dark solitons at the strong interaction of the first and second harmonics.

## 1. INTRODUCTION

The plasmons are excited in the conducting medium; the polaritons are excited as quasi-particles by interaction of photons and phonons in the dielectric medium [1]. The surface plasmon-polaritons (SPP) arise at the propagation of electromagnetic wave along the boundary of dielectric medium and metal as the mixture of photons, phonons and plasmons [2]. Recently the behaviors of SPP attract attention of researchers in connection with their unique properties for electronic technique at optical frequencies [3 - 6]. The SPP are "attached" on the boundary of media and are not radiated from the smooth boundary. The wavevectors of SPP have the values at several times greater then in air, i.e. less the wavelength of wave of the same frequency. The properties of nonlinear SPP modes, their types and interaction, the spatial solitons etc. were investigated in Refs. [2 - 5]. The dynamics of nonlinear polariton pulses in infinite dielectric media were analyzed in Refs. [6 - 8].

In our work we consider the nonlinear SPP pulses in the form of bright and dark solitons and analyze their generation at the interaction of the first and second harmonics TM-mode on the boundary of dielectric medium and nonmagnetic metal. The splitting of SPP in two fluxes that propagate with different velocities is proportional to the intensity of the first harmonic in quadratic polarization medium. Two fluxes of SPP are formed at the both harmonics of TM-mode on the boundary of media. We also obtain analytically and compute the shapes and transverse envelopes of nonlinear SPP pulses as bright and dark solitons at strong interaction of the first and second harmonics.

## 2. PLASMON-POLARITONS ON THE BOUNDARY OF DIELECTRIC MEDIUM AND METAL

We treat the mechanism of generation of the nonlinear SPP on the boundary of nonmagnetic dielectric medium with permittivity $e_1$ and magnetic permeability $m_1 = 1$ and nonmagnetic metal with $e_2$ and $m_2 = 1$ (Fig. 1).

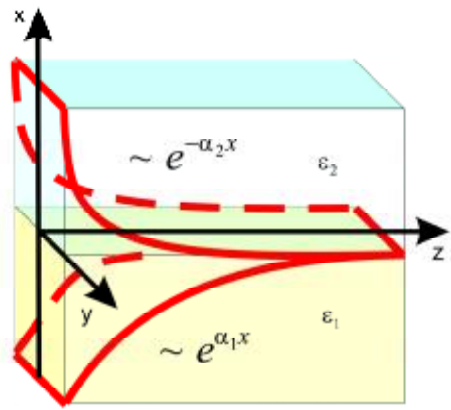

Fig. 1. The SPP on the boundary of dielectric medium and metal.

The three-dimensional wave falls on the inner boundary of dielectric medium and air angularly the total internal reflection and transforms to the surface wave. We suppose that the surface wave is exited as transverse magnetic (TM) mode with frequency $w$ and propagates on the dielectric and metal boundary along the axis z (Fig. 1). In the transverse plane the wave field damps at the both sides of boundary.

Suppose the initial electromagnetic wave is powerful, then we consider the nonlinear effects of the quadratic polarization of medium. In this case the SPP of TM-mode with transverse electric field polarization $E_x$ are formed at the first $\sim exp(-iwt)$ and second $\sim exp(-i2wt)$ harmonics due to the translation symmetry breaking on the boundary of mediums, and the SPP with longitudinal electric field polarization $E_z$ are formed only at the first harmonic $\sim exp(-iwt)$.

We represent the set of nonlinear equations for the SPP of TM-mode with components $E_x, E_z, B_y$ in the nonmagnetic dielectric medium

$$-\frac{\partial B_y}{\partial z} = \frac{e_1}{c}\frac{\partial E_x}{\partial t} + \frac{4p}{c}\frac{\partial}{\partial t}\left(c_{11}^{(2)}E_x^*E_x + c_{12}^{(2)}E_x^2\right), \quad \frac{\partial B_y}{\partial x} = \frac{e_1}{c}\frac{\partial E_z}{\partial t},$$
$$-\frac{\partial E_z}{\partial x} + \frac{\partial E_x}{\partial z} = -\frac{1}{c}\frac{\partial}{\partial t}B_y,$$
(1)

and in the nonmagnetic metal

$$-\frac{\partial B_y}{\partial z} = \frac{e_2}{c}\frac{\partial E_x}{\partial t} + \frac{4pg_2}{c}E_x, \quad \frac{\partial B_y}{\partial x} = \frac{e_2}{c}\frac{\partial E_z}{\partial t} + \frac{4pg_2}{c}E_z,$$
$$-\frac{\partial E_z}{\partial x} + \frac{\partial E_x}{\partial z} = -\frac{1}{c}\frac{\partial}{\partial t}B_y,$$
(2)

where $g_2 = iw_{2e}^2/4p(w + iG_2)$ is the conductivity of metal [9], $w_{2e}^2 = 4pe^2n_0/m$ is the electron plasma frequency, $G_2$ is the damping constant.

We consider that the wave field damps as $\sim exp(a_1 x)$ in the dielectric medium at moving off from the boundary to the subzero direction of axis $-x$, and it damps as $\sim exp(-a_2 x)$ in the metal at moving off from boundary to the positive direction of axis $x$, where $a_1 > 0, a_2 > 0$. The boundary conditions for TM-mode are $B_{y1} = B_{y2}$ and $E_{z1} = E_{z2}$ at $x = 0$, and we obtain the equation from the equation sets (1) and (2)

$$-\frac{e_1}{a_1} = \frac{e_2}{a_2} + i\frac{4pg_2}{a_2 w}.$$
(3)

The boundary condition for equation sets (1) and (2) is fulfilled in the case of the negative permittivity $e_2 < 0$, that is possible in the metal below plasma frequency [9].

## 3. DISPERSION EQUATION

In the linear case $c_{11}^{(2)} = 0$, $c_{12}^{(2)} = 0$ one can find the dispersion equation for SPP wave with components $E_{x,z}$, $B_y \sim exp(-iwt + ikz)$ in the dielectric medium (index 1) and in the metal (index 2) from the equation sets (1) and (2),

$$c^{-2}w^2 e_1 - k^2 + a_1^2 = 0, \tag{4}$$

$$c^{-2}w^2 e_2 - k^2 + a_2^2 + ic^{-2}4pwg_2 = 0. \tag{5}$$

We obtain the damping factors $a_1$ and $a_2$ from the equations (4) and (5) as

$$a_1 = \left(k^2 - c^{-2}w^2 e_1\right)^{1/2}, \qquad a_2 = \left(k^2 - c^{-2}w^2 e_2 - ic^{-2}4pwg_2\right)^{1/2}, \tag{6}$$

or calculate the wavevector $k$ by the Eq. (3) and Eq. (6) for the specified medium parameters $e_1$, $e_2$, $g_2$. The permittivity of dielectric medium is

$$e_1 = 1 + \frac{w_{1I}^2}{W_1^2 - w^2 - iG_1 w} + \frac{w_{1e}^2}{w_0^2 - w^2 - iG_1 w}, \tag{7}$$

where $w_{1e}^2 = 4p\, e^2 N_e m^{-1}$, $w_{1I}^2 = 4p\, e_{eff}^2 N_C m_{eff}^{-1}$ are the electron and ion plasma frequencies, $e_{eff}$, $m_{eff}$ are the effective charge and mass of ions, $W_1$ is the lattice resonance frequency, $w_0$ is the electron resonance frequency, $G_1$ is the damping constant. Considering both the ion and electron polarization of the metal crystal lattice under influence of the electromagnetic field we can renormalize its permittivity

$$e_2 = 1 + \frac{w_{2I}^2}{W_2^2 - w^2 - iG_2 w} - \frac{w_{2e}^2}{w^2 + iw_{re} w}, \tag{8}$$

where $W_2$ is the resonance frequency of metal lattice, $w_{re}$ is the frequency of plasmon relaxation in the metal. The second term in the Eq. (8) means the response of the ions of lattice, and the third term describes the response of electrons at electromagnetic field action. From the Eq. (8) it follows that the real part of the metal permittivity is less than zero $e_2 < 0$ for frequencies $w < w_{2e}$ at $w_I \ll w_e$ for the waves with frequencies far from the lattice resonance $W_2 \ll w$.

Substituting the damping factors $a_1$ and $a_2$ (6) into the Eq. (3) we obtain the dispersion equation for SPP of TM-mode

$$k^2 = \frac{w^2}{c^2} \frac{e_1 e_2}{e_1 + e_2},  \quad (9)$$

where $e_1$ and $e_2$ are defined by Eq. (7) and Eq. (8). Representing the dispersion equation (9) in the form of combined equations $k'^2 - k''^2 = \frac{w^2}{c^2}\tilde{e}'$, $k'k'' = \frac{w^2}{2c^2}\tilde{e}''$ we can found the real and imaginary parts of the wavevector $k' = \frac{w}{\sqrt{2}c}\left[(\tilde{e}'^2 + \tilde{e}''^2)^{1/2} + \tilde{e}'\right]^{1/2}$, $k'' = \frac{w}{\sqrt{2}c}\left[(\tilde{e}'^2 + \tilde{e}''^2)^{1/2} - \tilde{e}'\right]^{1/2}$, where $\tilde{e} = e_1 e_2 (e_1 + e_2)^{-1}$ is the effective complex permittivity, $\tilde{e} = \tilde{e}' + i\tilde{e}''$.

For example, the imaginary part of the wavevector $k'' = 2 \cdot 10^3 cm^{-1}$ at optical frequency $w = 3 \cdot 10^{15} s^{-1}$ for the glass with $e_1 = 4$ and silver with $e_2 = -10.7 + i0.3$ is small, and the real part $k' = 2.5 \cdot 10^5 cm^{-1}$ of SPP wavevector is greater than in air at 2.5 times, i.e. the SPP wavelength is less than in air at 2.5 times. Thereby we can obtain the wave at optical frequency with the subwave wavelength in this case.

## 4. NONLINEAR SPP AT THE FIRST AND SECOND HARMONICS

Consider the generation of nonlinear SPP at TM-mode by the powerful electromagnetic wave causing the nonlinear polarization of dielectric medium on the boundary of media. We can obtain the nonlinear equation for $E_x$ in the dielectric medium combining the equations of set (1)

$$\frac{\partial^2 E_x}{\partial z^2} - \frac{e_1}{c^2}\frac{\partial^2 E_x}{\partial t^2} + a_1^2 E_x$$
$$= \frac{4p}{c^2}\left(\frac{\partial^2}{\partial t^2} - \frac{a_1^2 c^2}{e_1}\right)\left(c_{11}^{(2)} E_x^* E_x + c_{12}^{(2)} E_x^2\right). \quad (10)$$

One can obtain other field components from the equation set (1) substituting the solution of Eq. (10).

We represent the components of electric field with transverse polarization in the form of the first and second harmonics $E_x = \tilde{E}_1(t,z)exp(-iwt + ik_1 z) + \tilde{E}_2(t,z)exp(-i2wt + ik_2 z)$, where $\tilde{E}_{1,2}(t,z)$ are the slowly varying amplitudes, $k_1 = c^{-1}w\sqrt{\tilde{e}(w)}$, $k_2 = 2c^{-1}w\sqrt{\tilde{e}(2w)}$. Equating the factors before exponents with identical arguments for the first and second harmonics and neglecting the higher order nonlinearities in the Eq. (10), we obtain combined equations for amplitudes of nonlinear SPP of TM-mode for the first and second harmonics

$$\frac{\partial \tilde{E}_1}{\partial z} + \frac{1}{v_1}\frac{\partial \tilde{E}_1}{\partial t} = ig_1 \tilde{E}_1^* \tilde{E}_2 \exp(iDkz),$$
$$\frac{\partial \tilde{E}_2}{\partial z} + \frac{1}{v_2}\frac{\partial \tilde{E}_2}{\partial t} = ig_2 \tilde{E}_1^2 \exp(-iDkz), \quad (11)$$

where $v_1 = c\dfrac{\sqrt{\tilde{e}(w)}}{e_1(w)}$, $v_2 = c\dfrac{\sqrt{\tilde{e}(2w)}}{e_1(2w)}$, $Dk = k_2 - 2k_1$, $g_1 = \dfrac{2p\,cc_{11}^{(2)}(w)}{w\sqrt{\tilde{e}(w)}}\left(\dfrac{a_1^2(w)}{e_1(w)} + \dfrac{w^2}{c^2}\right)$,

$g_2 = \dfrac{p\,cc_{12}^{(2)}(w)}{w\sqrt{\tilde{e}(2w)}}\left(\dfrac{a_1^2(w)}{e_1(w)} + \dfrac{4w^2}{c^2}\right)$.

## 5. WEAK INTERACTION OF HARMONICS

It is evident from the equation set (11) that the power of the first harmonic transforms to the power of the second harmonic and vice versa at the long length of interaction. Suppose the power of the first harmonic $I_1 = \tilde{E}_1^* \tilde{E}_1$ changes weakly, when transforming portion of power of the first harmonic to the second harmonic is negligibly small. In the weakly dispersive medium $Dk \approx 0$ we can represent the equation set (11) as

$$\left(\frac{\partial}{\partial z} + \frac{1}{v}\frac{\partial}{\partial t}\right)f_1 = i2g_1 I_1 f_2, \quad \left(\frac{\partial}{\partial z} + \frac{1}{v}\frac{\partial}{\partial t}\right)f_2 = ig_2 f_1, \quad (12)$$

where $f_1 = \tilde{E}_1^2$, $f_2 = \tilde{E}_2$, and obtain the partial solutions for the variables $f_1, f_2$ from equation set (12) as $f_1 = f_{10}\exp(-i2\tilde{w}t + i2\tilde{k}z)$, $f_2 = f_{20}\exp(-i2\tilde{w}t + i2\tilde{k}z)$, $f_{01,02} = const$, $v_1 \approx v_2 = v$. In this case the transverse electric fields of the first and second harmonics look like $\tilde{E}_1 = A_1\exp[-i(w+\tilde{w})t + i(k_1+\tilde{k})z]$, $\tilde{E}_2 = A_2\exp[-i2(w+\tilde{w})t + i(k_2+2\tilde{k})z]$, where $A_1 = \sqrt{f_{10}}$, $A_2 = f_{20}$. The relation of the phase additions $\tilde{w}$ and $\tilde{k}$

$$\tilde{k}_\pm = \tilde{w}v^{-1} \pm (g_1 g_2 I_1/2)^{1/2}, \quad (13)$$

where $\tilde{w}$ is the frequency of signal modulation. As it follows from the Eq. (13) the velocities of wave packets $\tilde{v}_\pm = Re[d(k+\tilde{k}_\pm)/dw]^{-1}$ for both SPP harmonics depend on intensity of the first harmonic $I_1$. Two fluxes of "slow" and "fast" nonlinear SPP are formed on the boundary of the media and propagate with different velocities, because the phase addition $\tilde{k}_\pm$ (13) has two values for all field components.

## 6. STRONG INTERACTION OF HARMONICS

If the intensity of the first harmonic is not constant $I_1 \neq const$, then it is possible to solve the equation set (11) by entering the real amplitudes and phases [10] as $\tilde{E}_{1,2}(t,z) = A_{1,2}(t,z) exp[if_{1,2}(t,z)]$ and separating the real and imaginary parts

$$\frac{\partial A_1}{\partial z} + \frac{v_1'}{|v_1|^2}\frac{\partial A_1}{\partial t} + \frac{v_1''}{|v_1|^2}\frac{\partial f_1}{\partial t}A_1$$
$$= -A_1 A_2 [g_1' sin(Dkz + f_2 - 2f_1) + g_1'' cos(Dkz + f_2 - 2f_1)],$$
$$\frac{\partial A_2}{\partial z} + \frac{v_2'}{|v_2|^2}\frac{\partial A_2}{\partial t} + \frac{v_2''}{|v_2|^2}\frac{\partial f_2}{\partial t}A_2$$
$$= A_1^2 [g_2' sin(Dkz + f_2 - 2f_1) - g_2'' cos(Dkz + f_2 - 2f_1)],$$
$$\frac{\partial f_1}{\partial z}A_1 + \frac{v_1'}{|v_1|^2}\frac{\partial f_1}{\partial t}A_1 - \frac{v_1''}{|v_1|^2}\frac{\partial A_1}{\partial t} \qquad (14)$$
$$= A_1 A_2 [g_1' cos(Dkz + f_2 - 2f_1) - g_1'' sin(Dkz + f_2 - 2f_1)],$$
$$\frac{\partial f_2}{\partial z}A_2 + \frac{v_2'}{|v_2|^2}\frac{\partial f_2}{\partial t}A_2 - \frac{v_2''}{|v_2|^2}\frac{\partial A_2}{\partial t}$$
$$= A_1^2 [g_2' cos(Dkz + f_2 - 2f_1) + g_2'' sin(Dkz + f_2 - 2f_1)],$$

where $v_{1,2} = v_{1,2}' + iv_{1,2}''$ are the complex velocities. The dynamics of amplitudes $A_1$ and $A_2$ is represented in Fig. 2 for the following boundary conditions: 1) the Gauss pulse $A_1|_{z=0} = exp(-t^2/T_0^2)$ is excited at the first harmonic and 2) the second harmonic has zero value $A_2|_{z=0} = 0$ at the input of medium.

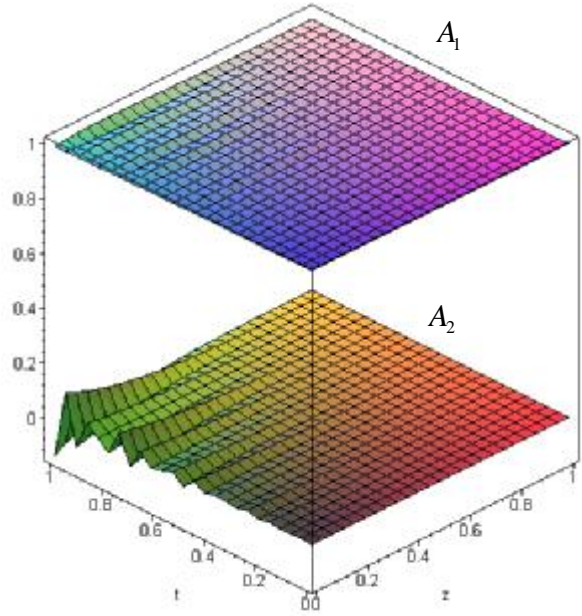

Fig. 2. The dynamics of SPP pulse amplitudes of the first $A_1$ and the second $A_2$ harmonics on the boundary of dielectric medium and metal.

We can transfer to the reference frame moving with pulse velocity $\tilde{v} \neq Re(v)$ by the introduction of variable $t = t - z/\tilde{v}$, and obtain the combined equations from the equation set (11) in the non-absorptive dielectric $v_{1,2}'' = 0, g_{1,2}'' = 0$,

$$\frac{dA_1}{dt} = -\tilde{g}_1 A_1 A_2 \sin F,$$
$$\frac{dA_2}{dt} = \tilde{g}_2 A_1^2 \sin F, \qquad (15)$$
$$\frac{dF}{dt} = -Dk\tilde{v} + \left(\tilde{g}_2 A_1^2 A_2^{-1} - 2\tilde{g}_1 A_2\right)\cos F,$$

where $F = Dkz + f_2 - 2f_1$ is the combination phase, $\tilde{g}_{1,2} = g_{1,2}' v_{1,2}' \tilde{v} / (\tilde{v} - v_{1,2}')$, $\tilde{v} > v_{1,2}'$. The equation set (15) has two motion integrals: 1) $I_{12} = \tilde{g}_1^{-1} A_1^2 + \tilde{g}_2^{-1} A_2^2$ and 2) $Dk\tilde{v} A_2^2 = 2\tilde{g}_2 A_1^2 A_2 \cos F$.

The combination phase can be constant $F = p/2$ due to the phase synchronism $Dk = 0$ in the weakly dispersive medium, because $\cos F = 0$ as it follows from the second integral of motion [10]. In this case we obtain the combined equations for amplitudes from the equation set (15)

$$\frac{dA_1}{dt} = -\tilde{g}_1 A_1 A_2, \quad \frac{dA_2}{dt} = \tilde{g}_2 A_1^2. \qquad (16)$$

We obtain the solutions for SPP pulse amplitudes at the first and second harmonics by the first integral of motion

$$A_1 = (\tilde{g}_1 I_{12})^{1/2} \cosh^{-1}\left[\tilde{g}_1(\tilde{g}_2 I_{12})^{1/2} t\right], \tag{17}$$

$$A_2 = (\tilde{g}_2 I_{12})^{1/2} \tanh\left[\tilde{g}_1(\tilde{g}_2 I_{12})^{1/2} t\right]. \tag{18}$$

The SPP pulse at the first harmonic (17) has the shape of bright soliton (Fig. 3, curve 2)

$$I_1 = \tilde{g}_1 I_{12} \cosh^{-2}\left[\tilde{g}_1(\tilde{g}_2 I_{12})^{1/2}(t - z/\tilde{v})\right], \tag{19}$$

and the pulse at the second harmonic (18) has the shape of dark soliton (Fig. 3, curve 3)

$$I_2 = \tilde{g}_2 I_{12} \tanh^2\left[\tilde{g}_1(\tilde{g}_2 I_{12})^{1/2}(t - z/\tilde{v})\right], \tag{20}$$

where $\tilde{v} = Re\{d[k + \tilde{k}_-(I_1)]/dw\}^{-1}$ is the pulse velocity depending on the first harmonic intensity $I_1$. The intensity of pulses at the first and second harmonics does not oscillate along the axis $z$ and time $t$ as the combination phase is constant due to the phase synchronism.

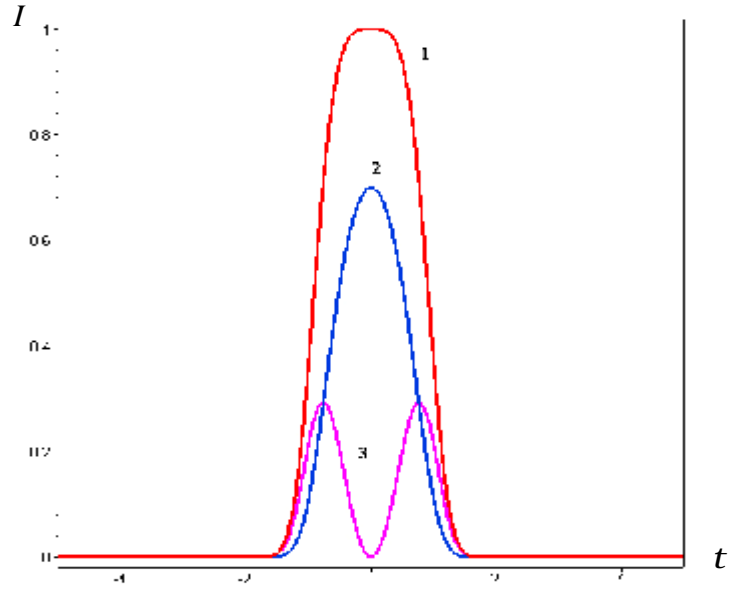

Fig. 3. The normalized intensities: the input Gauss pulse (curve 1) generates the pulses in the forms of bright soliton at the first harmonic (curve 2) and the dark soliton at the second harmonic (curve 3) if the combination phase is constant $F = p/2$.

## 7. CONCLUSION

The powerful three-dimensional incident wave transforms to the nonlinear SPP of the first and second harmonics of TM-mode, if the incident wave falls at the angle of total internal reflection on the boundary of dielectric medium and metal. "Slow" and "fast" fluxes of nonlinear SPP are formed on the boundary of the dielectric medium and metal and propagate with different velocities at the weak interaction of harmonics.

The incident wave can be modulated as a pulse train at the first harmonic. The pulses at the first harmonic excite the SPP pulses at the second harmonic. The SPP pulses can be formed as the bright and dark solitons at the strong interaction of the first and second harmonics.

These nonlinear effects arise in the optoelectronics devices and can be applied in design of new facilities based on the plasmon-polariton interactions [11, 12].

**REFERENCES**

1. Kittel C. *Quantum theory of solids*, New-York-London, John Wiley & Sons, 1963.
2. *Surface polaritons. Electromagnetic waves on surfaces and boundaries of mediums* / Edited by Agranovich V. M. and Mills D. L., Moscow, Nauka, 1985.
3. Davoyan A. R., Shadrivov I. V., Kivshar Y. S. "Quadratic phase matching in nonlinear plasmonic nanoscale waveguides", *Optics express*, **17**, No. 22 / 20063 (2009).
4. Davoyan A. R., Shadrivov I. V., Kivshar Y. S. "Self-focusing and spatial plasmon-polariton solitons", *Optics express*, **17**, No. 24 / 21732 (2009).
5. Lin Y.-Y., Lee R.-K., Kivshar Y. S. "Transverse instability of transverse-magnetic solitons and nonlinear surface plasmons", *Optics Letters*, **34**, No. 19, p. 2982-2984 (2009).
6. Dzedolik I. V., Lapayeva S. N., "Mass of polaritons in different dielectric media", *J. Opt.*, **13,** 015204 (2011).
7. Darmanyan S. A., Kamchatnov A. M., Neviere M. "Polariton effect in nonlinear pulse propagation", *Journal of Experimental and Theoretical Physics*, **96**, No. 5, p. 876-884 (2003).
8. Dzedolik I. V. *Polaritons in optical fibers and dielectric resonators*, Simferopol, DIP, 2007 (in Russian).
9. Shen Y. R. *The principles of nonlinear optics*, New-York, John Wiley & Sons, 1984.
10. Vinogradova M. B., Rudenko O. V, Sukhorukov A. P. *Theory of waves*, Moscow, Nauka, 1990 (in Russian).
11. Diniz L. O., Marega Jr. E., Frederico Nunes D., Borges B.-H. V., "A long-range surface plasmon-polariton waveguide ring resonator as a platform for (bio)sensor applications", *J. Opt.*, **13**, No. 11, 115001 (2011).

12. Zhao H., Li Y., Zhang G., "Study on the performance of bimetallic layer dielectric-loaded surface plasmon polariton waveguides", *J. Opt.*, **13**, No. 11, 115501 (2011).